\documentclass[aps,prd,showpacs,twocolumn,superscriptaddress,nofootinbib,preprintnumbers]{revtex4-1} 

\makeatletter
\def\p@subsection{}
\makeatother

\usepackage{color,graphicx,amsmath,amssymb,lipsum}
\usepackage[colorlinks=true,citecolor=blue,linkcolor=blue,urlcolor=blue, backref=false,pdfborder={0 0 0}]{hyperref}

\usepackage[normalem]{ulem}
\usepackage[utf8]{inputenc} 
\usepackage{mathtools}

\usepackage{float}

\usepackage{ulem}
\usepackage{diagbox}

\usepackage[dvipsnames]{xcolor}
\usepackage{mathrsfs}

\newcommand{\be}{\begin{equation}}
\newcommand{\ee}{\end{equation}}
\newcommand{\beqa}{\begin{eqnarray}}
\newcommand{\eeqa}{\end{eqnarray}}

\def\d{\partial}
\newcommand{\bseq}{\begin{subequations}}
\newcommand{\eseq}{\end{subequations}}

\renewcommand{\ln}{\mathop{\rm ln}\nolimits}

\def\gsim{\raise0.3ex\hbox{$\;>$\kern-0.75em\raise-1.1ex\hbox{$\sim\;$}}}
\def\lsim{\raise0.3ex\hbox{$\;<$\kern-0.75em\raise-1.1ex\hbox{$\sim\;$}}}

\def\beqn#1{\begin{equation}\label{#1}}
\def\eeqn{\end{equation}}

\def\beqa#1{\begin{eqnarray}\label{#1}}
\def\eeqa{\end{eqnarray}}

\def\Z2{$\mathcal{Z_2}$}

\newcommand {\ignore}[1]{}

\begin{document}


\title{
Vanishing of black hole
tidal Love numbers from 
scattering amplitudes
}

\author{Mikhail M. Ivanov}
\email{ivanov@ias.edu}
\affiliation{School of Natural Sciences, Institute for Advanced Study, 1 Einstein Drive, Princeton, NJ 08540, USA}
\author{Zihan Zhou}
\email{zihanz@princeton.edu}
\affiliation{Department of Physics, Princeton University, Princeton, NJ 08540, USA}

\begin{abstract} 
We extract the black hole (BH) static tidal deformability 
coefficients (Love numbers) and their spin-0 and spin-1
analogs by comparing on-shell amplitudes for fields 
to scatter off a spinning BH
in the worldline effective field theory (EFT)
and in general relativity (GR). 
We point out that the GR amplitudes
due to tidal effects
originate entirely from 
the BH potential region.
Thus, they can be separated from 
gravitational non-linearities in the wave region,
whose proper treatment requires 
higher order EFT loop calculations.
In particular, the elastic scattering 
in the near field approximation 
is produced 
exclusively by tidal effects.
We find this contribution 
to vanish identically, which implies that the static 
Love numbers of Kerr BHs
are zero
for all types of perturbations.
We also reproduce 
the known behavior of scalar Love numbers
for higher dimensional BHs.
Our results are manifestly gauge-invariant 
and coordinate-independent, 
thereby providing a valuable 
consistency check for the commonly used off-shell methods.
\end{abstract}

\maketitle

\textit{Introduction.---}
The worldline point-particle effective field theory 
for gravitational wave sources is a modern toolbox 
for precision waveform calculations~\cite{Goldberger:2004jt,Goldberger:2005cd,Porto:2016pyg,Goldberger:2022ebt}.
The increasing interest to these calculations 
is fueled by recent discoveries of
gravitational waves from black hole (BH) binaries~\cite{Abbott:2016blz}.
In the EFT each compact object in a binary 
is approximated as a point particle
at leading order, while the finite-size 
effects are captured by the higher-derivative
operators on the worldline. These operators
generate multipolar corrections to the 
point mass potential. The most general 
worldline theory for a perturbed spherically
symmetric body at leading order in derivatives of the long-wavelength
metric field is given by
\be
\label{eq:fseft}
S_{\rm pp}=-m\int d\tau 
+c_E\int d\tau E_{\mu\nu}E^{\mu\nu}
+c_B\int d\tau B_{\mu\nu}B^{\mu\nu}\,,
\ee
where $m$ is the mass of a compact body, 
$E_{\mu\nu}$ and $B_{\mu\nu}$
are the electric (parity even)
and magnetic (parity odd)
parts of the Weyl tensor.\footnote{We work in the unit system
$G=c=\hbar=1$.} 
The dimensionfull 
Wilson 
coefficients $c_{E,B}$ 
measure the gravitational response 
of the body to external tidal
fields 
in the quadrupolar 
sector. 
A simple response calculation shows
that 
the action~\eqref{eq:fseft}
generates the following
Newtonian potential $\Phi_N$~\cite{Kol:2011vg,Cardoso:2019vof,Hui:2020xxx,Charalambous:2021mea} (schematically), 
\be
\Phi_N =  \frac{m}{r}-\sum_{m=-2}^2\mathcal{E}_{2m}r^2\left(1-\frac{c_E}{r^5}\right)\,,
\ee
where 
$\mathcal{E}_{2m}$ are tidal harmonic coefficients, 
we assumed a long distance limit 
and ignored the magnetic part for simplicity. 
This expression coincides with a classic Newtonian
definition of the static tidal response,
which is controlled by the 
static deformability coefficients, ``Love numbers''~\cite{PoissonWill2014,Binnington:2009bb}.
Thus, $c_{E,B}$ provide us with a gauge-invariant 
definition of Love numbers in general relativity (GR). 
In order to extract the Love numbers
for a particular object, one needs to match 
some quantity calculated in the full theory (GR)
and the EFT.

The Love numbers capture
details of body's internal structure,
e.g. they depend on the equation of state 
for neutron stars~\cite{Flanagan:2007ix}.
For BHs, however, the application 
of the EFT leads to a surprising result.
Matching the EFT and GR calculations 
for BH static perturbations 
implies that their Love numbers vanish identically
in GR in 
four dimensions~\cite{Damour:2009vw,Binnington:2009bb,Kol:2011vg,Hui:2020xxx,Chia:2020yla,Charalambous:2021mea}.
This is in sharp contrast with the 
dimensional analysis 
(Wilsonian naturalness), suggesting that $c_{E,B}\sim r_s^{5}$, where $r_s$ is the Scwarzschild radius.
The vanishing of 
the EFT Wilson coefficients
for BHs has been a major puzzle in BH physics for many years~\cite{Porto:2016zng}, see~\cite{Charalambous:2021kcz,Hui:2021vcv,Hui:2022vbh,Charalambous:2022rre} 
for 
proposals addressing this problem.

The vanishing of black hole Love numbers 
is however, still a controversial topic. 
The problem is that this result is obtained
by comparing certain static field profiles
in GR with the corresponding EFT 
off-shell one-point functions
that are calculated in a particular coordinate system.
In theory, this should not be a problem
as one is free to choose any quantity  
for matching and the result should not depend
on whether this quantity is gauge-invariant or not. 
In practice, however, the matching 
in specific gauges may be difficult, 
and often leads to conflicting conclusions, 
see e.g. a recent debate in the literature 
on whether the Love 
numbers vanish for Kerr black holes~\cite{Poisson:2014gka,LeTiec:2020spy,Chia:2020yla,Charalambous:2021mea},
and discussions on interference between the 
tidal effects and Post-Newtonian corrections~\cite{Kol:2011vg,Poisson:2020vap,Charalambous:2021mea,Creci:2021rkz,
Ivanov:2022hlo}.
This leaves 
room for doubt
about results of the off-shell Love 
number extraction, see e.g.~\cite{Gralla:2017djj,Kim:2021rfj}.

To avoid any confusion, it is desirable 
to match the EFT and GR 
by comparing manifestly gauge invariant quantities, 
such as amplitudes to elastically scatter 
off a BH geometry in GR versus an S-matrix element in the EFT.
One can easily estimate that the Love 
operator contributes to the
graviton-BH 
cross-section at order $\sim \omega^8c_E^2\sim \omega^8r_s^{10}$.
Since the EFT is valid
when $r_s\omega \ll 1$,
this term is much smaller
than the leading long-range (Newtonian)
contribution $\sim r_s^2$, and also than 
relativistic corrections to that contribution.
The total cross-section
should be of the form, schematically
\be
\label{eq:fullcr}
\sigma_{\rm GR}(\omega) = 
\underbrace{r_s^2}_{\text{Newton}}
\underbrace{(1+(r_s\omega)^2+...)}_{\text{rel. corrections}}
+\underbrace{r_s^{10}\omega^8}_{\text{finite size}}\,.
\ee
The smallness of the finite-size contribution
makes it hard to extract. 
At face value, one needs to compute the full 
GR cross-section 
to the fifth Post-Minkowski (PM) order
in 
black hole perturbation theory (BHPT).
In addition, one needs to 
calculate the EFT amplitude at the same 5PM,
which is a four loop 
calculation in the EFT nomenclature,\footnote{Strictly speaking, the leading finite-size contribution in~\eqref{eq:fullcr}
is a 3PM term that comes from the interference between the Love number  
and the Newtonian amplitudes. This, however, will be irrelevant for our
further discussion.
}
see~\cite{Kalin:2020mvi,Kalin:2020fhe,Cheung:2020sdj,Bern:2020uwk,Bern:2021dqo} for recent progress in these calculations, 
including tidal effects.

In this \textit{Letter} we 
show that in fact, there is a consistent 
approximation 
to BHPT, where the total cross-section 
is given exclusively by the finite-size 
operators. 
Using this approximation, the 
matching with the EFT can be performed entirely 
at the tree level.

\textit{General EFT for Kerr black holes.---}
A general astrophysical 
relevant BH is described by the Kerr 
metric~\cite{Kerr:1963ud}.
The Kerr black hole has two horizon, $r_{\pm}=M\pm \sqrt{M^2-a^2}$, where $M$ and $a$ denote BH mass and spin.
Within the EFT, the worldline action inherits
symmetries of the underlying gravitational 
background, i.e. the EFT for spinning black holes 
must have axial symmetry. 
In the context of 
static tides 
it is sufficient
to incorporate this fact by promoting the 
Love numbers in~\eqref{eq:fseft} 
to tensors~\cite{Goldberger:2020fot,Charalambous:2021mea}, i.e. 
considering the following EFT for 
metric perturbations, 
\be
\label{eq:quadkerr}
S_{\text{finite size}}=
\int d\tau ~ \lambda^{i'j'}_{ij}E^{ij}E_{i'j'} +\text{magnetic}\,,
\ee
where we have switched to the BH rest frame, and $\lambda^{i'j'}_{ij}$ is a symmetric real matrix. 
The above local EFT coupling only captures 
conservative effects. In order to account for
dissipation we introduce the coupling between 
the long-wavelength tidal field and 
the composite mass quadrupole $Q_{ij}(X)$~\cite{Goldberger:2005cd,Goldberger:2020fot,Ivanov:2022hlo}, 
\be
\label{eq:kerrdiss}
S_{\rm diss}=\int d\tau~ Q_{ij}(X) E^{ij} + \text{magnetic}\,,
\ee
where $X$ are gapless degrees of freedom on the horizon.
Actions~\eqref{eq:quadkerr} and \eqref{eq:kerrdiss}
can be used to compute, e.g. a potential 
contribution to the Weyl curvature
scalar for $\ell=2$~\cite{Charalambous:2021mea} (schematically),
\be
\label{eq:p0}
\begin{split}
& \psi^{(\ell=2)}_{0} \sim
\sum_{m=-2}^2{}_{2} Y_{\ell m}(\theta,\phi)
\mathcal{E}_{2 m} \left(1-\frac{k^{(2)}_{2 m}+i\nu^{(2)}_{2 m}}{
r^{5}}\right)\,, 
\end{split}
\ee
where ${}_{2} Y_{\ell m}$ is 
the spin-weighted spherical harmonic, 
$k^{(2)}_{2 m}$
is a real coefficient
related to $\lambda_{ij, i'j'}$,
whist $\nu^{(2)}_{2 m}$
is a real number 
generated by the time-reversal
odd (non-conservative) 
part of 
the retarded correlator 
$\langle Q_{ij} Q_{i'j'}\rangle_{\text{ret}}$~\cite{Charalambous:2021mea}.
The factor in front of the $r^{-5}$ term in 
Eq.~\eqref{eq:p0} is called the response coefficient. 
Note that its real part 
captures the conservative effects (Love numbers),
while the imaginary part is responsible for 
BH absorption (dissipation numbers)~\cite{Chia:2020yla,Charalambous:2021mea}.

The local action~\eqref{eq:quadkerr}
can be generalized to the case of a generic test field with
an angular multipole $\ell$ and a positive
integer spin $s$.
The most general leading order EFT action 
needed to reproduce the
effect of static Love numbers 
and their spin-0 and spin-1 analogs
is given by
\be
\label{eq:geneft}
\begin{split}
  \sum_{\ell=s}\frac{1}{2\ell!}\int d\tau  \Bigg[&\lambda^{(0)}{}^{L}_{L'} \partial_{\langle L \rangle} \phi 
\partial^{\langle L' \rangle} \phi \\
&+ \lambda^{(1)}{}^{L}_{L'} \partial_{\langle L {-1} } E_{i_\ell \rangle}
\partial^{\langle L' {-1} } E^{i'_\ell \rangle} 
\\
&+ \lambda^{(2)}{}^{L}_{L
'} \partial_{\langle L {-2} } 
E_{i_{\ell-1} i_\ell \rangle}
\partial^{\langle L' {-2} } E^{i'_{\ell-1} i'_\ell \rangle}  \Bigg] \,,
\end{split}
\ee
where we have omitted the magnetic contributions for brevity,
$L=i_1...i_\ell$ is the multi-index, 
$\d_L\equiv \d_{i_1}...\d_{i_\ell}$,
$\langle L \rangle$ denotes the trace-free part,
$E^i$ and $\phi$ are test electric and scalar fields.

The Wilson tensors $\lambda^{(s)}{}^{L'}_L$
have previously been extracted by comparing 
the static EFT off-shell  one-point
functions like \eqref{eq:p0}
with BHPT calculations, which implied 
that all such tensors vanish identically~\cite{Chia:2020yla,Charalambous:2021mea}.
In contrast, the dissipation numbers 
$\nu^{(2)}_{\ell m}$ we found to be non-zero~\cite{LeTiec:2020spy}.
These results, however, were obtained in a 
coordinate-dependent fashion. Now we show how 
the Love numbers and dissipation numbers
can be extracted from the on-shell scattering 
and absorption cross-sections 
in a gauge-invariant manner.

\textit{Scattering off a Kerr black hole in general relativity.---}
The scattering by rotating black holes is an old 
and well studied subject~\cite{1978ApJS...36..451M,1988sfbh.book.....F,Dolan:2008kf}. 
We present
only few essential elements in this \textit{Letter} 
and leave
other 
details for future work~\cite{future}. 
Consider a wave of a spin-$s$ test field 
impinging along the axis of a Kerr BH.
The scattering cross-section is given by
\be
\frac{d\sigma}{d\Omega}=|f_s(\theta)|^2 + 
|g_s(\theta)|^2 \,,
\ee
where $\theta$ is the scattering angle, $f_s(\theta)$ and $g_s(\theta)$
are helicity-conserving
and helicity-reversing amplitudes, respectively. 
Their partial wave expressions are given by 
\begin{widetext}
\be
\left\{
    \begin{array}{ll}
      f_s(\theta) \\
      g_s(\theta)   
    \end{array}
  \right\} 
  = \frac{\pi}{i\omega}
  \sum_{P=\pm 1}\sum_{\ell =s}^\infty 
  \left\{
  \begin{array}{ll}
      {}_{-s}S^s_\ell(\theta,a\omega)\\
      {}_{-s}S^s_\ell(\pi - \theta,a\omega)P(-1)^{\ell + s}
    \end{array}
  \right\} 
   {}_{-s}S^s_\ell(0,a\omega)[\eta_{\ell s}^P\exp(2i\delta^P_{\ell s})-1]\,,
\ee
\end{widetext} where $P=\pm 1$ is parity,
${}_{-s}S^m_\ell(\theta,a\omega)$
are spin-weighted spheroidal harmonics~\cite{Teukolsky:1972my,Teukolsky:1973ha},
$\exp(2i\delta^P_{\ell s})$
is the 
scattering phase, $\eta_{\ell s}^P$ is the 
transmission factor that captures absorption by the BH horizon.

The transmission coefficients 
and scattering phases are extracted 
from a separable equation for 
perturbations around Kerr black holes
known as the Teukolsky equation~\cite{Teukolsky:1972my,Teukolsky:1973ha}.
Each type of perturbations is encapsulated 
in a specific 
Newman-Penrose (NP) scalar
that can be used to compute a corresponding energy flux.
In order to compute the scattering cross-section 
we need the NP scalars with spin weights $(-s)$,
which factorize as 
\[ 
\psi_{-s} = e^{im\phi}e^{-i\omega t}   
{}_{-s}S_{ \ell}^{m}(\theta; a\omega) 
{}_{-s}R_{\ell m}(r)\,.\]
For the physically
relevant solution, the radial 
part ${}_{-s}R_{\ell m}$
satisfies the purely in-going
boundary condition at the BH horizon
and
has the following asymptotic at spatial infinity
($r_*$ is the tortoise coordinate),
\be
 {}_{-s}R_{ \ell m}\sim 
 \underbrace{B^{\text{(inc)}}_{-s\ell m} r^{-1}e^{-i\omega r_*}}_{\text{ingoing wave}}
 + 
 \underbrace{B^{\text{(refl)}}_{-s \ell m} r^{2s-1}e^{i\omega r_*}
 }_{\text{outgoing wave}}\,,
\ee
where $B^{\text{(inc)}}_{-s \ell m}$
and $B^{\text{(refl)}}_{-s \ell m}$
are complex constants.
The phase shifts are given by
\be
\eta^P_{\ell s} e^{2i\delta^P_{\ell s}} 
= (-1)^{\ell + 1} 
\frac{A_s}{(2\omega)^{2s}}
\frac{B^{\text{(refl)}}_{-s \ell s}}{
	B^{\text{(inc)}}_{-s \ell s}}\,,
\ee
where $A_s$ are normalization factors
related to the Starobinsky-Teukolsky constants~\cite{future}.

The constants $B^{\text{(inc)}}_{-s \ell s}$
and $B^{\text{(refl)}}_{-s \ell s}$ should be 
extracted from the solution of the appropriate
Tekolsky equations. 
Mano, Suzuki, and Takasugi (MST) have constructed such
solutions in a systematic low-frequency expansion
~\cite{Mano:1996vt,Mano:1996gn}.
The solution takes the form of an infinite 
series over
hypergeometric
functions in the near zone (potential) region,
and of an infinite series of  
Coulomb wavefunctions in the far zone (wave) region. 
$B^{\text{(inc)}}_{-s \ell m}$
and $B^{\text{(refl)}}_{-s \ell m}$ are determined 
by matching these two series. 
The MST solution is a linear combination of 
modes labeled by the ``renormalized''
angular momentum $\nu = \ell + O(\epsilon^2)$,
where $\epsilon\equiv 2M\omega$ is the PM expansion parameter, 
and $\ell \geq s$ is the usual integer 
angular momentum number.

The partial wave amplitude in the sector $\nu$ and $s$,
formally to all powers in
frequency, 
is given by
\begin{widetext}
\be
\label{eq:MST}
\begin{split}
& \frac{B^{\text{(refl)}}_{s \ell m}}{B^{\text{(inc)}}_{s \ell m}}=
\frac{1}{\omega^{2s}}
\textcolor{blue}{\underbrace{\frac{
	1+ie^{i\pi\nu}\frac{K_{-\nu-1}}{K_\nu}}
	{1-ie^{-i\pi\nu}\frac{\sin(\pi(\nu-s+i\epsilon))}{\sin(\pi(\nu+s-i\epsilon))}\frac{K_{-\nu-1}}{K_\nu}}}_{
	\text{Near zone}} }
	\textcolor{red}{
	\underbrace{\times \frac{A_-^\nu}{A_+^\nu}
	e^{i\epsilon(2\ln \epsilon -({1-\kappa}))}}_{
	\text{Far zone}}}\,,
\end{split} 
\ee
\end{widetext}
where $\kappa = \sqrt{1-a^2/M^2}$,
and $A_+$, $A_-,~K_\nu,~K_{-\nu-1}$ are some $\epsilon$-dependent 
coefficients.
Crucially, the above expression factorizes into 
two distinctive contributions. 
The first term (in blue) comes from 
the matching between the potential and 
the wave region expansions. It is sourced by 
the perturbed gravitational potential 
in the near zone and hence 
contains information 
about finite-size effects encoded
in the multipole expansion
of BH perturbations. 
The second term (in red) stems from the solution 
of the Teukolsky equation in the wave zone.
By construction, it contains PM terms due to non-linearity of 
gravity but no information about the BH finite size structure.
Crucially, this separation holds only
in linear BHPT.

 \begin{figure*}
    \centering    \includegraphics[width=0.8\textwidth]{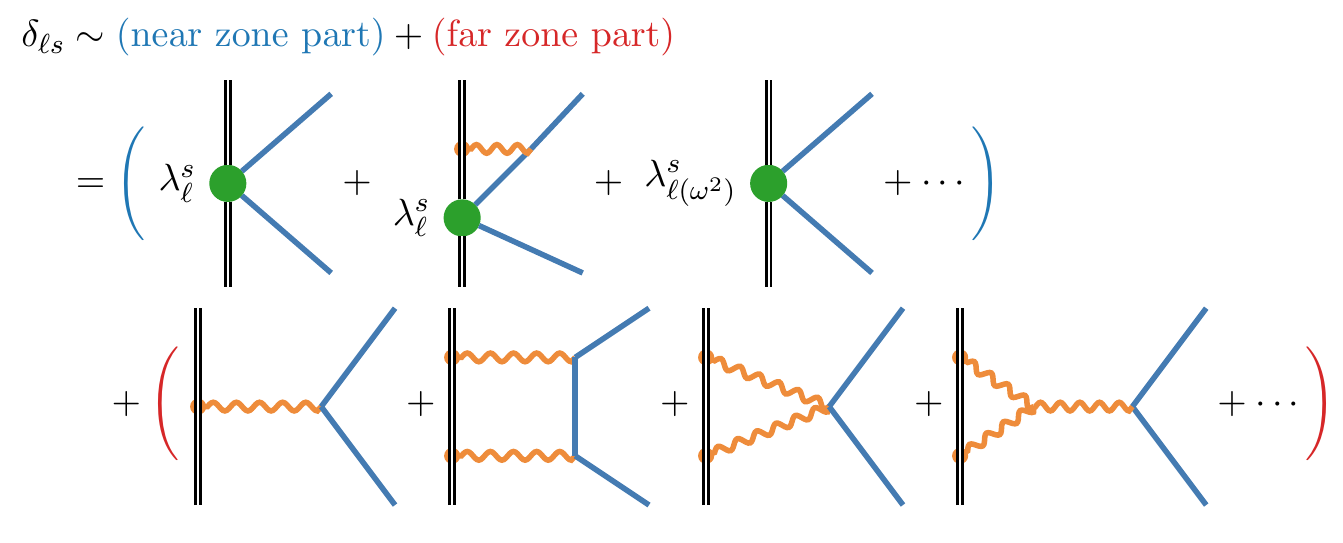}
    \caption{Diagrammatic interpretation of the elastic scattering amplitude in the EFT and its representation
    in terms of the near zone and far zone parts
    in GR. Curly lines stand for off-shell gravitons.
    The vertical straight lines depict the worldline.
    External legs correspond to external spin-$s$ on-shell fields.
    The upper diagrams stem from the EFT finite-size
    action. The lower diagrams 
    describe the scattering of the on-shell
    fields off background
    potential modes in the far region.}
    \label{fig:diags}
\end{figure*}

The interpretation of Eq.~\eqref{eq:MST} 
within the EFT is not straightforward. 
We may associate 
the two terms in Eq.~\eqref{eq:MST} with two  sets of EFT diagrams shown 
in Fig ~\ref{fig:diags}.
The wave zone term captures the scattering of 
on-shell gravitons off the BH geometry. 
Naively, it is this part that maps onto the EFT PM loop
expansion.
Indeed, if we ignore the potential zone contribution
and retain the 1PM corrections 
from the radiation zone, we recover 
the Newtonian answer, e.g.
\be 
\frac{d\sigma^{\rm GR}}{d\Omega}\Big|_{\text{1PM}}
=\frac{M^2}{\sin^4\left(\frac{\theta}{2}\right)}
\left(
\cos^8\left(\frac{\theta}{2}\right) 
+ \sin^8\left(\frac{\theta}{2}\right)
\right)\,,
\ee 
for a Schwarzschild BH and for the spin-2 field.
This result can be reproduced with the tree-level
EFT diagrams analogous to the one shown in Fig.~\ref{fig:diags}
~\cite{Bautista:2021wfy,Saketh:2022wap,future}.
In contrast, 
the near field contribution in Eq.~\eqref{eq:MST},
physically, 
incorporates finite-size effects.
Specifically, the first order near zone expansion 
should capture the static Love number contribution.
The relativistic $\epsilon$-
corrections to this result 
correspond to
the graviton-dressed Love number diagrams and frequency-dependent local worldline operators
$\sim \lambda_{\ell(\omega^2)} \dot E^2_{ij}$ (``dynamical Love numbers'')~\cite{Charalambous:2021mea,Ivanov:2022hlo},
shown in Fig.~\ref{fig:diags}.
These appear starting at second order
in the near zone expansion. 

The above arguments suggest that the separation
between the near zone and far zone contributions 
should hold in the EFT. 
This may not be true if EFT loop diagrams 
produce logarithmic divergences that
should be renormalized by the Love number.\footnote{Here we employ dimensional regularization, where non-logarithmic 
divergence vanish identically.} 
The logarithms then indicate
the mixing between loops and finite-size 
operators in
the EFT, which complicates the matching 
with the GR results.\footnote{For 
general renormalized angular momentum 
$\nu$ there are no logarithms in the near zone part of 
GR amplitudes, 
and hence there is 
no ambiguity. The logarithmic 
corrections appear only in the limit $\nu \to \ell\in \mathbb{N}$. This is akin to 
dimensional regularization. }
Strictly speaking, a full calculation is needed in this
case. However, since logarithms should be present both in the EFT 
and UV calculations, one may identify 
the scheme-independent (running) part of the Love numbers
from the GR solution. In what follows we will see
that for four dimensional Kerr BH scattering there
are no logs and thus no mixing, and hence the matching 
between the EFT and GR is unambiguous.

\textit{Scattering in the near zone approximation.---}
Let us neglect the PM 
terms completely and compute 
the cross-section entirely from the 
near zone term. 
This amounts to using 
an approximate BHPT solution 
obtained through the leading order
matching 
of the potential 
and radiation regions~\cite{1973JETP...37...28S,1974JETP...38....1S,Page:1976df,Chia:2020yla}\footnote{This solution can be recovered as the $\epsilon\to 0$ limit
of Eq.~\eqref{eq:MST}.}.
At this order one matches the first order
near zone solution with the 
zeroth order far zone solution (describing 
a free motion with an angular momentum $\ell$)
in an overlapping region where both solutions are valid.
Note that this approximation is 
unacceptable 
from the practical point of view as it 
misses the leading order contributions
in~\eqref{eq:fullcr}. 
However, it is perfectly suitable 
for our goal 
to extract the finite size effects.

In the leading near zone approximation the parity-even
and parity-odd phase shifts take the same
expressions and hence the helicity 
reversing amplitude vanishes.
Assuming that the scattering is perturbative,
the  
total cross-section
in the partial wave approach
can be decomposed
into the elastic and absorptive 
contributions:
\be
\label{eq:eabs}
\begin{split}
&   \sigma_{\rm elastic} = \frac{4\pi}{\omega^2} \sum_{\ell = s}^\infty (2\ell + 1)\sin^2\delta_{\ell s}\,,\\
& \sigma_{\rm abs} = \frac{\pi}{\omega^2} \sum_{\ell = s}^\infty (2\ell + 1) \Big(1 - \eta_{\ell s}^2 \Big)\,.
\end{split} 
\ee
Performing a matching calculation, we get
\be
\label{eq:NZres}
 \begin{split}
 &\eta_{\ell s} e^{2i\delta_{\ell s}} 
 =\\
 &1 + i (-1)^s \frac{(\ell + s)! (\ell - s)!}{(2\ell) ! (2\ell + 1)!} \Big(2 \omega ( r_+ -  r_-)\Big)^{2\ell + 1} \mathcal{I}_{-s \ell s} 
\,,
 \end{split}
\ee
where $\mathcal{I}_{s \ell m}$ is the harmonic 
near zone
response function~\cite{Chia:2020yla,Charalambous:2021mea},
\begin{equation}\label{4D Response Function for General Spin}
    \mathcal{I}_{s \ell m} 
    = i(-1)^{s+1} P_{+} \frac{(\ell+s) !(\ell-s) !}{(2 \ell) !(2 \ell+1) !} \prod_{j=1}^{\ell}\left(j^2+4 P_{+}^2\right)\,,
\end{equation}
and  $P_+  \equiv \frac{a m-2 {r}_+ {\omega}}{r_+-r_-}$.
With Eq.~\eqref{eq:NZres}, we get
\begin{equation}
    \begin{aligned}
        \eta_{\ell s} &= 1 - (-1)^s \frac{(\ell + s)! (\ell - s)!}{(2\ell)! (2\ell + 1)!} \Big(2  \omega ( r_+ -  r_-)\Big)^{2\ell + 1} {\rm Im} \mathcal{I}_{-s \ell s} ~, \\
        \delta_{\ell s} &= \frac{1}{2} (-1)^s \frac{(\ell + s)! (\ell - s)!}{(2\ell) ! (2\ell + 1)!} \Big(2  \omega ( r_+ -  r_-)\Big)^{2\ell + 1} {\rm Re} \mathcal{I}_{-s \ell s} ~.
    \end{aligned}
\end{equation}
Since ${\rm Re} \mathcal{I}_{-s \ell s}=0$
for Kerr BHs, we
conclude that the scattering 
phase shift is zero for all spins $s$ and multipoles $\ell$,
to all orders of the BH spin,
\be
\sigma^{\rm GR}_{\rm elastic, \ell} = 0 \,.
\ee
Comparing this with the EFT result 
from the local action~\eqref{eq:geneft} we conclude 
that all static 
Kerr Love numbers vanish identically, 
in agreement with previous off-shell
calculations~\cite{Chia:2020yla,Charalambous:2021mea}.

As far as absorption
is concerned, 
it is straightforward to see from Eq.~\eqref{eq:eabs} that
\be 
\begin{split}
    \sigma_{\rm abs, \ell} =&
\frac{2 (-1)^s\pi}{\omega^2} (2\ell + 1)  \frac{(\ell + s)! (\ell - s)!}{(2\ell)! (2\ell + 1)!}  \\
& \times \Big(2 \omega (r_+ -  r_-)\Big)^{2\ell + 1} {\rm Im} 
\mathcal{I}_{-s \ell s}\,.
\end{split}
\ee
This generalizes 
the result of~\cite{Page:1976df}
for the $\ell=s$ case.
Note that the imaginary part of the response coefficient 
generates 
the absorption cross-section and 
is directly linked with the EFT dissipation numbers~\eqref{eq:p0}~\cite{Chia:2020yla,Goldberger:2020fot,Charalambous:2021mea,Ivanov:2022hlo}.

\textit{Love numbers of higher dimensional BHs.---}
Love numbers do not vanish
in general  
for BHs in spacetimes with a number of dimensions $d$ greater than four~\cite{Kol:2011vg,Hui:2020xxx}. 
Let us perform their 
explicit matching from near zone scattering amplitudes.
For simplicity, we focus 
on scalar fluctuations of 
higher dimensional static BHs.
The corresponding phase shift is given by
\be 
  \eta_\ell e^{2 i \delta_\ell} =
1 + i 
\frac{
2^{1-\hat{d} - 2\ell} 
\pi (r_+ \omega)^{\hat d + 2\ell}
}
{\Gamma( \frac{\hat d}{2} + \ell) \Gamma(1 + \frac{\hat d}{2} + \ell)} 
\mathcal{I}_\ell \,,
\ee
where $\hat d \equiv d-3$, $\hat \ell\equiv \ell/\hat d$, and the response function 
\begin{equation}
\label{eq:highd}
    \mathcal{I}_\ell = 
    \frac{\Gamma(-2\hat \ell - 1) \Gamma(1+ \hat \ell) \Gamma(1 + \hat \ell - \frac{2 i  r_+ \omega}{\hat d})}{\Gamma(-\hat \ell) \Gamma(2\hat \ell + 1)\Gamma(-\hat \ell - \frac{2 i r_+ \omega}{\hat d})} ~.
\end{equation}
From this equation, we get
\begin{equation}\label{Higher Dimensional Schwarzschild Spin-0 Phase Shift}
    \begin{aligned}
        \eta_\ell & = 1 - \frac{2^{1-\hat d - 2\ell} \pi (r_+ \omega)^{\hat d + 2\ell}}{\Gamma( \frac{\hat d}{2} + \ell ) \Gamma (1 + \frac{\hat d}{2} + \ell )} {\rm Im} \mathcal{I}_\ell\,, \\
        \delta_\ell & = \frac{2^{- \hat d - 2\ell} \pi (r_+ \omega)^{\hat d + 2\ell}}{\Gamma( \frac{\hat d}{2} + \ell) \Gamma(1 + \frac{\hat d}{2} + \ell)} {\rm Re} \mathcal{I}_\ell ~. 
    \end{aligned}
\end{equation}
On the EFT side, the worldline action
\begin{equation}
    S_{\text{finite size}} =  \int d\tau \sum_{\ell = 0}^\infty \frac{\lambda_\ell^{s=0}}{2 \ell!} \Big( \partial_{\langle L \rangle} \phi \Big)^2 
\end{equation}
yields the following tree-level scattering amplitude 
in the orbital sector $\ell$,
\be
\begin{split}
iT 
& = i \frac{\lambda_{\ell}^{s=0}}{\ell !} \omega^{2\ell} \frac{\ell ! (\hat d - 2)!!}{(2\ell + \hat d -2)!!} \frac{(\ell + \hat d -1)!}{\ell ! (\hat d -1)!} P_{\ell}^{(d)}(\cos\theta) \,,
\end{split}
\ee
where $P_{\ell}^{(d)}$
are generalized Legendre polynomials.
Comparing this with 
the phase shift from the UV theory~\eqref{Higher Dimensional Schwarzschild Spin-0 Phase Shift}, we obtain
\be 
    \lambda_\ell^{s=0}=(-1)^\ell \frac{\pi^{\frac{d-1}{2}}}{2^{\ell-2}} \frac{\Gamma\left(\frac{5-d}{2}-\ell\right)}{\Gamma\left(\frac{5-d}{2}\right) \Gamma\left(\frac{d-3}{2}\right)} r_+^{2 \ell+d-3} {\rm Re}\mathcal{I}_\ell \,,
\ee
which identically coincides with the 
result of off-shell matching from~\cite{Hui:2020xxx}.
Note that according to Eq.\eqref{eq:highd} 
the Love numbers vanish if $\hat\ell$ is an integer,
exhibit the classical renormalization group running
if $\hat\ell$ is half integer, and  
are constant numbers $O(r_+^{2\ell+d-3})$ otherwise,
in full agreement with the off-shell results~\cite{Kol:2011vg,Hui:2020xxx}.

\textit{Conclusions.---}
We have matched on-shell amplitudes 
of massless fields scattering off BHs in the 
worldline EFT and the near zone approximation of BHPT. 
We have found 
that the static tidal Love numbers
vanish for four dimensional 
spinning (Kerr) black holes to all orders of the black hole spin. 
Our results are gauge-invariant,
and hence they remove any uncertainty as to the
validity of the previous results based 
on the coordinate-dependent off-shell 
matching. 

We have also reproduced the known behavior of the spin-0 Love numbers of BHs in a general number
of dimensions. 
This is an important consistency 
check of our approach and of the former off-shell calculations.
We stress that the mapping between the EFT 
and GR potential and wave regions does not 
work if Love numbers run  logarithmically.
Nevertheless, 
we can still extract the scheme-independent 
part from the full GR solution in this case~\cite{future}.

Two facts are key to the matching of 
elastic cross-sections: the realization
that finite-size effects 
originate from the BH near region, and the 
absence of logarithms.
These have allowed us to bypass both a full 
calculation of four loop 
corrections in the EFT, and 
the construction of the Teukolsky equation solution
at the 5PM order.
It would be interesting to extend our results to the time-dependent local worldline couplings
(``dynamical Love numbers''), 
which are non-zero for four dimensional Kerr black holes~\cite{Charalambous:2021mea}.
These correspond to a second order 
near zone approximation.
It is also important 
to understand the implications of the Love symmetry
for on-shell observables.
Indeed, 
the Love symmetry explains the vanishing 
of Love numbers in the off-shell calculations
as a result of an algebraic constraint~\cite{Charalambous:2021kcz}. 
It would be curious to see if this constraint 
manifests itself
at the level of on-shell scattering amplitudes.

\textit{Acknowledgments.}
We thank Horng Sheng Chia, 
Sergei Dubovsky, and
Gregor K{\"a}lin  
for helpful discussions.
We are grateful to Panos Charalambous, 
Alfredo Guevara,
Lam Hui, Austin Joyce, 
Julio Parra-Martinez
for their comments on the draft.
We are especially grateful to 
Rafael Porto for detailed feedback on the draft.  

\bibliography{short.bib}

\end{document}